\documentclass[aps, prl, reprint, groupedaddress, longbibliography]{revtex4-1}
\usepackage{graphicx}
\graphicspath{ {Figures/} }
\usepackage{natbib}
\usepackage{amsmath,amssymb}
\usepackage{subfig}
\usepackage{setspace}
\usepackage{float}

\usepackage[usenames, dvipsnames]{color}
\usepackage{hyperref}
\hypersetup{
	hyperindex,
	breaklinks,
	colorlinks=true,
	linkcolor=blue,
	citecolor=magenta,
	bookmarks=true,
	bookmarksopen=true,
	bookmarksopenlevel=2,
	pdfstartpage={1},
	pdfstartview={FitH},
	pdfview={FitH 0},%pdfstartview=FitH,pdfview=FitH,%pdfstartview={XYZ null null 1},
	pdfauthor={B. F. Farrell and P. J. Ioannou},
	pdftitle={}}

\usepackage{ifthen}

\def\U{\bm{\mathsf{U}}}
\def\Uv{\mathbf{U}}

\def\U{\bm{\mathsf{U}}}

\def\Uv{\boldsymbol{U}}

\newcommand{\be}{\begin{equation}}
\newcommand{\ee}{\end{equation}}
\newcommand{\bdm}{\begin{equation*}}
\newcommand{\edm}{\end{equation*}}
\newcommand{\bea}{\begin{eqnarray}}
\newcommand{\eea}{\end{eqnarray}}

\newcommand{\partialf}[2]
{
 \ifthenelse{\equal{#1}{}}{\frac{\partial}{\partial #2}}{\frac{\partial #1}{\partial #2}}
}

\renewcommand{\(}{\left(}
\renewcommand{\)}{\right)}

\providecommand\bnabla{\boldsymbol{\nabla}}
\providecommand\bcdot{\boldsymbol{\cdot}}

\newcounter{saveeqn}%

\def\Uv{\mathbf{U}}

\def\av{\mathbf{a}}

\newcommand{\defn}{\ensuremath{\stackrel{\mathrm{def}}{=}}}
\renewcommand{\equiv}{\defn}

\providecommand\bnabla{\boldsymbol{\nabla}}
\providecommand\bcdot{\boldsymbol{\cdot}}

\renewcommand{\Re}{Re}

\renewcommand{\U}{\mathbf{U}}
\renewcommand{\u}{\mathbf{u}}

\begin{document}

\title{Statistical State Dynamics Based Study of the Role of Nonlinearity in the Maintenance of Turbulence in Couette Flow}

% Force line breaks with \\
%\title{Self-sustaining Turbulence in the Restricted Nonlinear Systems}% Force line breaks with \\

\author{Brian F. Farrell}
\affiliation{Department of Earth and Planetary Sciences, Harvard University}
\author{Petros J. Ioannou}
\email{pjioannou@phys.uoa.gr}
\affiliation{Department of Physics, National and Kapodistrian University of Athens}
\author{Marios-Andreas~Nikolaidis}
\affiliation{Department of Physics, National and Kapodistrian University of Athens}
%\affiliation{Department of Mechanical Engineering, Johns Hopkins University}
%\author{Vaughan Thomas}
%\affiliation{Department of Mechanical Engineering, Johns Hopkins University}
%

\date{\today}% It is always \today, today,

\begin{abstract}
While linear non-normality underlies the mechanism of energy transfer from the externally driven flow to the perturbation field that sustains turbulence, nonlinearity is also known to play an essential role. The goal of this study is to better understand the role of nonlinearity in sustaining turbulence. The method used in this study is implementation in Couette flow of a statistical state dynamics (SSD) closure at second order in a cumulant expansion of the Navier-Stokes equations in which the averaging operator is the streamwise mean. The perturbations in this SSD are the deviations from the streamwise mean and two mechanisms potentially contributing to maintaining these second cumulant perturbations are identified. These are parametric perturbation growth arising from interaction of the perturbations with the fluctuating mean flow and transient growth of perturbations arising from nonlinear interaction between components of the perturbation field. By the method of comparing the turbulence maintained in the SSD and in the associated direct numerical simulation (DNS) in which these mechanisms have been selectively included and excluded, parametric growth is found to maintain the perturbation field of the turbulence while the more commonly invoked mechanism associated with transient growth of perturbations arising from scattering by nonlinear interaction is found to suppress perturbation growth. In addition to verifying that the parametric mechanism maintains the perturbations in DNS it is also verified that the Lyapunov vectors are the structures that dominate the perturbation energy and energetics in DNS. It is further verified that these vectors are responsible for maintaining the roll circulation that underlies the self-sustaining process (SSP) and in particular the maintenance of the fluctuating streak that supports the parametric perturbation growth.
 \end{abstract}


\maketitle

%% SECTION NAMES:
%% Should be lower-case except for proper nouns and abbreviations
%% Should not end with a period
%=========================================================================
%=========================================================================

\section{Introduction}

Turbulence is widely regarded as the primary exemplar of an essentially nonlinear 
phenomenon. However, the mechanism by which energy is transferred in shear flows from
the externally forced component of the flow to the broad spectrum of spatially and 
temporally varying perturbations is through linear non-normal interaction between these 
components \cite{ Boberg-Brosa-1988,Henningson-Reddy-1994,Farrell-Ioannou-1994b,Kim-Lim-2000,Farrell-Ioannou-2012}. 
Nevertheless, nonlinearity participates in an essential way in the
cooperative interaction between the mean and the perturbation
 by which turbulence self-sustains. Our goal in this study is to provide a
more comprehensive understanding of the role of nonlinearity and its interaction with linear
non-normality in the maintenance of turbulence. 

Because realistic wall-turbulence is maintained by the statistical state dynamics (SSD) of the Navier-Stokes
equations closed at second order with the averaging operator chosen to be the streamwise
mean \cite{Farrell-Ioannou-2012,Farrell-etal-2016-PTRSA,
Farrell-Ioannou-2017-sync} it is inviting to study the mechanism of wall-turbulence using this SSD which
has the advantage of complete analytic characterization. We employ SSD-based analysis to
examine the role of nonlinearity in turbulence maintenance, specifically its role in maintaining  the perturbations 
from the streamwise mean at statistical equilibrium.

A commonly invoked physical 
process by which this maintenance of the perturbation field is hypothesized to occur is through
recycling of perturbations which have completed their 
transient amplification 
into new perturbations that are at the initial stage 
of transient growth leading to
renewed growth and in this way to turbulence 
maintenance \cite{Boberg-Brosa-1988,Trefethen-etal-1993, Gebhardt-Grossmann-1994,Baggett-Trefethen-1997, 
Grossmann-2000}.  This idea underlies  the regeneration cycle which was inspired by observations
in which streak breakdown 
produces perturbations configured to give rise to
new streak formation \cite{Jimenez-1994}.
%as well as a number of toy models of turbulence dynamics \cite{Trefethen-etal-1993, Gebhardt-Grossmann-1994,Baggett-Trefethen-1997, 
%Grossmann-2000}. 
An alternative mechanism of perturbation maintenance that has been shown to support turbulence
in the second order SSD of
a variety of turbulent shear flows is a process of parametric growth in which fluctuation of
the streamwise mean flow maintains the perturbation 
field \cite{Farrell-Ioannou-2012,Thomas-etal-2014,Farrell-etal-2016-PTRSA,
Farrell-etal-2016-VLSM,Farrell-Ioannou-2017-sync}.

%In this work we assess
%the relative contributions of the parametric growth process and the perturbation-perturbation 
%nonlinearity  process in maintaining the perturbation variance.  
%Because a  realistic turbulence in Couette flow 
%is maintained by  second order closure of the Navier-Stokes equations in which the 
%averaging operator is chosen to be the streamwise mean it is 
%inviting to examine the mechanism of turbulence maintenance in shear flow  
%using the associated statistical state dynamics (SSD) as a platform  \cite{Farrell-Ioannou-2012,Farrell-etal-2016-PTRSA,
%Farrell-Ioannou-2017-sync}.  

%Moreover, %being a second order closure and consequently quasi-linear \cite{Herring-1963}, 
%the SSD we use supports turbulence on a highly restricted component of the full 
%nonlinearity of the  Navier-Stokes equations; specifically, the nonlinearity 
%between perturbations from the streamwise mean 
%flow is not retained and the success of this SSD in supporting the turbulence shows
%that this nonlinearity and therefore the regeneration mechanism previously described is not 
%necessary for maintaining turbulence  \cite{Farrell-Ioannou-2012,Bretheim-etal-2015,Farrell-etal-2016-VLSM}.    

%in which  the perturbation-perturbation nonlinearity 
%is either neglected  or retained 
%in order to 
%examine in detail the mechanism maintaining the perturbation component of the turbulence. 
%The relative contribution of these two nonlinear mechanisms 
%can be examined through these second order SSDs. 
It was shown previously that 
realistic turbulence is  maintained by restricting the SSD dynamics to the second of these 
mechanisms; this was done by simply neglecting the perturbation-perturbation 
nonlinearity in the second order 
closure \cite{Farrell-Ioannou-2012,Thomas-etal-2015,Bretheim-etal-2015,Farrell-etal-2016-VLSM}.
However, although  these results establish that
the perturbation-perturbation nonlinearity is not necessary for perturbations from the streamwise mean as well as the turbulence itself 
to be maintained in a second order closure, the influence of perturbation-perturbation 
nonlinearity on the turbulence is still of interest because
it has been implicated  in Navier-Stokes turbulence dynamics by interpretations of DNS data 
\cite{Jimenez-1994,Hamilton-etal-1995,Jimenez-Pinelli-1999,Kim-Lim-2000}
and also in part because perturbation-perturbation 
nonlinearity alone has been  shown to maintain turbulent state analogues in 
simple model  systems \cite{Trefethen-etal-1993, Gebhardt-Grossmann-1994,Baggett-Trefethen-1997,
Grossmann-2000}.
Given that the parametric mechanism  supports realistic turbulence in the absence of perturbation-perturbation
nonlinearity,  the experiment available to us is to include the perturbation-perturbation
nonlinearity and assess  the influence of the addition of this term  on the parametrically maintained
turbulence.

%The two nonlinear 
%mechanisms described are isolated by these SSD  
%perturbation growth arising from parametric instability of the
%fluctuating mean flow and  transient perturbation growth of perturbations 
%produced by scattering associated with perturbation-perturbation nonlinearity.  

%A commonly invoked physical 
%process in which pertubation-pertubation nonlinearity sustains turbulence
%is that in which  perturbation nonlinearity is supposed to recycle perturbations which have completed their 
%transient amplification into new perturbations that are at the initial stage of transient growth leading to
%renewed growth and in this way to turbulence 
%maintenance.  This idea underlies  the regeneration cycle which was inspired by observations
%in which streak breakdown 
%produces perturbations configured to give rise to
%new streak formation by the linear non-normal lift-up mechanism \cite{Jimenez-1994}
%as well as a number of toy models of turbulence dynamics (references). 

%In this work approximations to the SSD of Couette flow turbulence at $R=600$ are 
%examined in order to assess the relative importance
%of parametric growth and perturbation-perturbation nonlinearity in the maintenance of this turbulence.  
The specific  SSD model examined is a reduced non-linear  
model  (RNL) in which the second cumulant  is approximated as the  state covariances obtained from the perturbation dynamics 
in which the nonlinearity has been neglected. We find by comparing RNL simulations made using this model
with DNS that the turbulence and its energetics are similar whether the
nonlinear interactions between perturbations from the streamwise mean are retained or neglected.
This result demonstrates that  the parametric 
mechanism dominates in the maintenance of the turbulence
and in fact closer examination of the energetics reveals that 
the perturbation-perturbation nonlinearity rather than serving 
to support the turbulence actually decreases effectiveness of energy transfer from the mean
  to the perturbations.  Furthermore, it is also verified that the parametrically 
  maintained Lyapunov vectors analytically predicted to support the turbulence by RNL 
  that dominate the perturbation
energy and energetics in DNS.  It  is further verified 
  that moreover these vectors are also found to be responsible for maintaining the roll circulation 
  that underlies the self sustaining process (SSP) and in particular the maintenance of 
  the fluctuating streak that supports the parametric perturbation growth.

 %This paper is structured as follows: after formulation of the equations in section 2, we calculate the Lyapunov exponents

\section{Formulation}

In order to study the mechanism by which nonlinearity between
streamwise varying components in a turbulent shear flow participate in
the maintenance of turbulence 
we begin
by partitioning  the velocity field of  plane parallel Couette flow into streamwise mean and perturbation components,
or equivalently  into the $k_x=0$ and the $k_x \ne 0$ components
of the  Fourier decomposition of the flow field, where $k_x$ is the
wavenumber in the streamwise, $x$, direction. In this decomposition the flow field is partitioned as:
\begin{equation}
\u = \Uv(y,z,t) + \u'(x,y,z,t)~,
\label{eq:mean}
\end{equation}
with cross-stream direction $y$ and spanwise direction  $z$. It is important to note
that in this decomposition the mean flow retains
temporal variation in its spanwise structure
and particularly that this mean flow includes the time-dependent streaks.
The mean  used in the cumulant expansion is 
fundamental to formulating an SSD that retains the physical
mechanism of turbulence in shear flow.   
 The centrality of spanwise variation of the mean flow, which is associated with 
the fluctuating streak component, to the maintenance of turbulence
has been  demonstrated by numerical experiments that show
turbulence is not sustained when the streaks are sufficiently
damped or removed~\cite{Jimenez-Pinelli-1999}.
Given that in  the SSD turbulent state 
spanwise and temporal inhomogeneity are required to allow the parametric
 instability of the  fluctuating streamwise streak to be supported it is necessary 
to allow both spanwise and temporal variations in the mean operator used to 
define the cumulants in the SSD
in \eqref{eq:mean}.% in order to analyze the turbulent state.

The non-dimensional Navier-Stokes equations expressed using this mean and perturbation partition are:
\begin{subequations}\label{eq:NS}
\begin{gather}
\partial_t\U + \underbrace{\U  \bcdot \bnabla  \U }_{N_1}   + \bnabla  P -  \Delta \U /R = \underbrace{- \langle\u ' \bcdot \bnabla  \u '\rangle_x}_{N_2}\ ,
\label{eq:NSm}\\
 \partial_t\u '+  \underbrace{ \U  \bcdot \bnabla  \u ' +
\u ' \bcdot \bnabla  \U }_{N_3}  + \bnabla   p' -  \Delta  \u '/R\nonumber\\
= \underbrace{- \( \u ' \bcdot \bnabla  \u ' - \langle\u ' \bcdot \bnabla  \u '\rangle_x \,\)}_{N_4} ~,%+ \sqrt{\varepsilon} ~{\mathbf{f}}'(x,y,z,t)\ ,
 \label{eq:NSp}\\
 \bnabla  \bcdot \U  = 0\ ,\ \ \ \bnabla  \bcdot \u ' = 0\ ,%\ \ \ \bnabla  \bcdot \mathbf{f}' = 0\,
 \label{eq:NSdiv0}
\end{gather}\label{eq:NSE0}\end{subequations}
where $R= U_w h/ \nu$ is the Reynolds number and $\pm U_w$ the wall velocity at $y=\pm h$.
The  flow satisfies no-slip boundary conditions in the cross-stream direction: $\U (x,\pm h,z,t)= (\pm U_w,0,0)$,
$\u '(x,\pm h,z,t)=(0,0,0)$ and periodic boundary conditions  in the $z$ and $x$ directions. 
Lengths are nondimensionalized by $h$,  velocities by $U_w$, and time by
 $h/U_w$. Averaging is denoted with angle brackets $\langle \bcdot \rangle$ with 
 the bracket subscript indicating the averaging variable, so that e.g. the streamwise 
 mean velocity is $\Uv \equiv \langle \u \rangle_x = L_x^{-1} \int_0^{L_x} \u ~dx$, 
 where $L_x$ is the streamwise length of the channel.
The Navier-Stokes equations
with this decomposition are referred to as the DNS system.   In \eqref{eq:NS} we have indicated 
with an underbrace the nonlinear terms in  DNS  of primary relevance to our study. In the 
streamwise mean flow equation~\eqref{eq:NSm}  nonlinear interactions among
$k_x=0$ flow components are referred to as  $N_1$ and  Reynolds stress divergence term  produced by 
nonlinear interaction between the $k_x$ and $-k_x$ flow components
 with $k_x \ne 0$, is referred to as $N_2$.  In the perturbation equation~\eqref{eq:NSp}  the
interaction
 between the instantaneous streamwise  mean flow and  
 the $k_x \ne0$ flow components is referred to as $N_3$.
 If the mean flow $\U$ is a solution of~\eqref{eq:NS},
 interaction $N_3$ between the perturbations and this mean flow $\U$  can be viewed in the perturbation equation 
 \eqref{eq:NSp}  as a linear 
 interaction. From that perspective, in the perturbation equation ~\eqref{eq:NSp}
 transfer of energy from the mean to the perturbations is due to linear 
 non-normal interaction between $\U$ and the perturbations, $\u'$, although from the 
 perspective of the DNS system \eqref{eq:NS} as a whole this term is nonlinear.  
 This is a crucial point in the analysis to follow as we will be taking $\U$ in $N_3$ to be known and this 
 term to be linear from the perspective of the perturbation equation \eqref{eq:NSp}.  Finally, the nonlinear interaction
  between perturbation components $k_{x_1}\ne 0$ and  $k_{x_2} \ne 0$, with $k_{x_1}\ne -k_{x_2}$
 is referred to as $N_4$.

Transition to and maintenance of a self-sustained turbulent state results
even when only nonlinearity $N_2$ and term $N_3$ are retained~\cite{Farrell-Ioannou-2012}. By retaining both 
nonlinearities $N_1$ and $N_2$ and term $N_3$
we obtain the restricted non-linear system (RNL):
\begin{subequations}
\label{eq:RNS}
\begin{gather}
\partial_t\U + \U  \bcdot \bnabla  \U   + \bnabla  P -  \Delta \U /R = - \langle\u ' \bcdot \bnabla  \u '\rangle_x\ ,
\label{eq:RNSm}\\
 \partial_t\u '+   \U  \bcdot \bnabla  \u ' +
\u ' \bcdot \bnabla  \U   + \bnabla   p' -  \Delta  \u '/R
= 0 ~,%+ \sqrt{\varepsilon} ~{\mathbf{f}}'(x,y,z,t)\ ,
 \label{eq:RNSp}\\
 \bnabla  \bcdot \U  = 0\ ,\ \ \ \bnabla  \bcdot \u ' = 0\ .%\ \ \ \bnabla  \bcdot \mathbf{f}' = 0\,
 \label{eq:RNSdiv0}
\end{gather}\label{eq:RNSE0}
\end{subequations}
%Using the velocity non-divergence $p'$ in~\eqref{eq:RNSp} is determined to be  a bilinear function of $\U$ and $\u'$
%and can be  eliminated with~\eqref{eq:RNSp} written in the symbolic form:
%\begin{equation}
% \partial_t\u ' = \A(\U) \u'~,
%%+ \sqrt{\varepsilon} ~{\mathbf{f}}'(x,y,z,t)\ ,
% \label{eq:NSfa}
%\end{equation}
%indicating that if we assume that we have obtained the function $\U(y,z,t)$, the perturbation velocity $\u'$ is consistently obtained
%by solving~\eqref{eq:NSfa} which governs the evolution of  linear perturbation on mean flow $\U$.
It has been confirmed that this RNL system supports a realistic self-sustaining process (SSP) which maintains a turbulent state in
minimal channels~\cite{Farrell-Ioannou-2012,Farrell-Ioannou-2017-bifur}, in channels of moderate sizes at  both low and high Reynolds numbers (at least for $R_\tau \le 1000$)
\cite{Thomas-etal-2014,Farrell-etal-2016-VLSM,Farrell-etal-2016-PTRSA},  and also
in very long channels~\cite{Thomas-etal-2015}.
%The maintenance of a realistic  turbulent state as $L_x$ increases in RNL turbulence argues that turbulence could be maintained with
%rectified streaks of infinite length. In that sense the
%chosen decomposition identifies the proper physical components of the dynamics underlying the SSP that sustains the turbulent state.
%The question we address in this note  is what is the role of nonlinearity N_4 in NL turbulence.

Consider in isolation the time varying mean flow $\U$ obtained from a state of  turbulence 
either of the RNL  or the DNS system.
Sufficiently small perturbations, $\u'$,  to this mean flow evolve  according  to~
\begin{equation}
 \partial_t\u '+  \U  \bcdot \bnabla  \u ' +
\u ' \bcdot \bnabla  \U   + \bnabla   p' -  \Delta  \u '/R~=0~,~~\bnabla  \bcdot \u'=0~,
%+ \sqrt{\varepsilon} ~{\mathbf{f}}'(x,y,z,t)\ ,
 \label{eq:RNSp1}
\end{equation}
which is  the perturbation equation~\eqref{eq:RNSp} of the RNL system, 
while in the DNS system the finite perturations $ \u'$ obey
the different equation~\eqref{eq:NSp} with the $N_4$ term included. 
In the  self-sustained RNL turbulence the perturbations, $\u'$, that evolve
under the linear dynamics~\eqref{eq:RNSp} or equivalently under 
\eqref{eq:RNSp1}
remain finite and bounded. Therefore 
the mean-flow, $\U$, of the  RNL turbulent state  is stable in the sense that perturbations, 
i.e. the streamwise varying flow  components, $\u'$,
that evolve under~\eqref{eq:RNSp1},  have  zero asymptotic growth rate and  the mean flow
can be considered to be in the critical state of neutrality, poised between stability and instability. 
A question that will be addressed in this paper is  whether  the mean flow, $\U$, 
that is obtained from a DNS  shares this property of being adjusted 
similarly to neutrality in the sense 
that perturbations,  $\u'$,
that evolve under~\eqref{eq:RNSp1}, remain bounded and therefore have  vanishing 
asymptotic growth rate and  the mean flow of the DNS
can therefore be 
considered to be similarly in a
 critical state of parametric neutrality when proper account is taken of dissipation.
If the turbulent mean flow  $\U$ of the DNS can be shown to be neutral,
in this sense of parametric neutrality, and the associated perturbations can 
be shown to be the Lyapunov vectors of this $\U$,
then  the  mechanism of turbulence identified 
analytically in the RNL system, in which the  perturbations arise from parametric instability of the
mean flow  with the mean flow being regulated to neutrality through quasi-linear interaction with the
perturbation field,  will have been extended to DNS.  Identification of DNS
turbulence dynamics with that of RNL would represent a 
fundamental advance in understanding because RNL 
turbulence  is fully and analytically characterized so that this identification 
would imply extension of the full analytical characterization of 
RNL turbulence to the DNS system.
 For this program to succeed it is required to show that the dynamically 
 substantive difference between the RNL system and the DNS, which 
 is the appearance in DNS of the perturbation-perturbation nonlinearity 
 $N_4$, does not fundamentally change the dynamics of turbulence.

An illustrative aspect of the insight that can be gained by identifying in the DNS system
the mechanisms that are known to 
be operating in the RNL system relates to the adjustment of turbulence to a statistically 
stationary state.  The mechanism by which
the statistical state of turbulence in the RNL 
system  is regulated to its statistical mean
can be related to an influential conjecture 
that in a turbulent system the linear instability of the mean state is adjusted 
by quasi-linear interaction with the perturbations to a state of modal neutrality \cite{Malkus-1956, Herring-1963,Stone-1978}.
The theory of turbulence based on the second order SSD we use has
in common with this influential hypothesis
the concept of adjustment by quasi-linear interaction 
between the mean flow and perturbations to neutrality as the general mechanism 
determining the statistical state of turbulence.  
%
%
%
%Consistent with the adjustment to neutrality conjecture we
%have  shown that  RNL turbulence is equilibrated to its statistical 
%equilibrium by a variant of this concept of adjustment to 
%neutrality. 
%We differ  with previous work in the particulars of this program however, including 
%considering the parametric instability of the time 
%dependent streamwise mean state to be the instability that is
%adjusted to neutrality rather than inflectional instability of the
% time, spanwise and streamwise mean flow.
While turbulent convection~\citep{Malkus-1954, Malkus-Veronis-1958} 
%and the baroclinic turbulence in the midlatitude atmosphere~\citep{Stone-1978}
displays a usefully close approximate adherence
to modal neutrality  when both the spatial and temporal means 
are taken to define the mean flow,  the turbulent mean state of wall-bounded
shear flows, defined as the streamwise, spanwise
and temporal mean, $\langle \U \rangle_{z,t}$,
is  hydrodynamically stable and far from neutrality in apparently strong violation of
the adjustment to neutrality conjecture~\cite{Reynolds-Tiederman-1967}.
However, study of RNL turbulence suggests that this program is
essentially correct and can be extended to wall-turbulence
requiring only the additional recognition that the instability  to be equilibrated is the instability of the
time-dependent operator associated
with linearization about the temporally varying streamwise mean flow. Among the theoretical advances 
arising from identifying the mechanism of RNL turbulence and that of 
DNS is extension of this physical mechanism determining the statistical steady state to DNS turbulence.

The maximum growth  rate of perturbations to the streamwise 
mean  governed by the linear dynamics of~\eqref{eq:RNSp1} is given by 
the  top Lyapunov exponent
of $\u'$  defined as:
\begin{equation}
\lambda_{Lyap} = \lim_{t \to \infty} \frac{\log |\u'|}{t}~.
\label{eq:exp}
\end{equation}
RNL turbulence with  mean~\eqref{eq:mean}  satisfies
the neutrality  conjecture  precisely under our interpretation because for RNL
\begin{equation}
\lambda_{Lyap} = 0 ~.
\label{eq:exp0}
\end{equation}

An issue we wish to examine in this work is  whether DNS turbulence (with
the $N_4$ term included) is
similarly neutral in the Lyapunov sense with its perturbations being 
supported by parametric 
growth on its fluctuating mean flow and with perturbation structure being  that 
predicted by the associated 
Lyapunov vector structures.
Specifically,  whether fluctuations, $\u'$, evolving 
under~\eqref{eq:RNSp1}  on the  time dependent mean flow, $\U$, 
that has been obtained from a turbulent DNS,
 have $\lambda_{Lyap}$, as defined in~\eqref{eq:exp}, zero 
when proper account is taken of dissipative processes  and in addition whether the
predicted Lyapunov structure can be verified to be maintaining the perturbations in the DNS.
We caution the reader 
that the Lyapunov structures and  exponents we are calculating  are not  the
familiar Lyapunov structures and  exponents associated with small perturbations to the full turbulent
state trajectory.  This more familiar use of Lyapunov exponents and associated 
vectors is concerned with  growth  of
perturbations $\delta \U$, $\delta \u'$  to the tangent linear
dynamics of the full Navier-Stokes equations linearized about the entire turbulent
trajectory $\U$, $\u'$. This tangent linear dynamics
typically has many positive Lyapunov 
exponents \cite{Keefe-etal-1992}. We instead  
calculate the Lyapunov exponents and structures only of perturbations, $\u'$, evolving
under the linear dynamics~\eqref{eq:RNSp1} about the  time dependent  mean flow $\U$
and there are typically only a small set of these that correspond to the perturbation component of the turbulent state which 
are neutrally stable while the rest are damped.
It is also important to recognize that 
the parametric perturbation evolution equation~\eqref{eq:RNSp1} governing
 the perturbation dynamics of RNL 
 is not limited to small perturbation amplitude because the 
 perturbation equation is strictly linear  and the nonlinearity required to regulate the perturbations to their finite statistical 
 equilibrium state is not explicit in~\eqref{eq:RNSp1} but rather is contained
  in  the Reynolds stress feedback term $N_2$ appearing in the mean equation
  which serves to provide feedback regulation of the mean state to neutral Lyapunov stability.

The mean  used in the cumulant expansion is 
fundamental to formulating an SSD that retains the physical
mechanism of turbulence in shear flow.   
 The centrality of spanwise variation of the mean flow, which is associated with 
the streak component, to the maintenance of turbulence
has been  demonstrated by numerical experiments that show
turbulence is not sustained when the streaks are sufficiently
damped or removed~\cite{Jimenez-Pinelli-1999}.
Given that in  the SSD turbulent state 
spanwise and temporal inhomogeneity are required to allow the parametric
 instability of the the fluctuating streamwise streak to be supported it is necessary 
%for dynamic consistency  with the broken symmetries of the SSD
to allow both spanwise and temporal variations in the mean operator used to 
define the cumulants in the SSD
which requires representation~\eqref{eq:mean}. The requirement that the streamwise 
mean be taken to support  turbulence in a second order SSD 
provides a partition of the mechanisms by which nonlinearity 
enters the dynamics.   This naturally compelled partition is into the 
completely characterized nonlinearity mechanism of the RNL 
dynamics and the remaining nonlinearity that has not yet been 
completely characterized which is that  contained in the $N_4$ term of the 
DNS system perturbation equation.

% in order to analyze the turbulent state.

%It  remains an open question  whether in the turbulent state the streamwise statistical homogeneity is also broken.
%The evidence that is usually presented to support  breaking of the  streamwise statistical homogeneity
%is that no arbitrarily long streamwise  structures  are observed in simulations, but
%this evidence is based on individual realizations which may not  reflect the symmetries of the statistical state.
%The evidence supporting the non-breaking of the streamwise symmetry is that RNL 
%has been shown to support a realistic turbulent state under the assumption
%of a streamwise mean ($k_x=0$) in long turbulent channels,
% demonstrating  that there exists  a statistical turbulent state with statistical streamwise homogeneity~\citep{Thomas-etal-2015}. 
% 

\begin{figure*}
\centering\includegraphics[width=24pc]{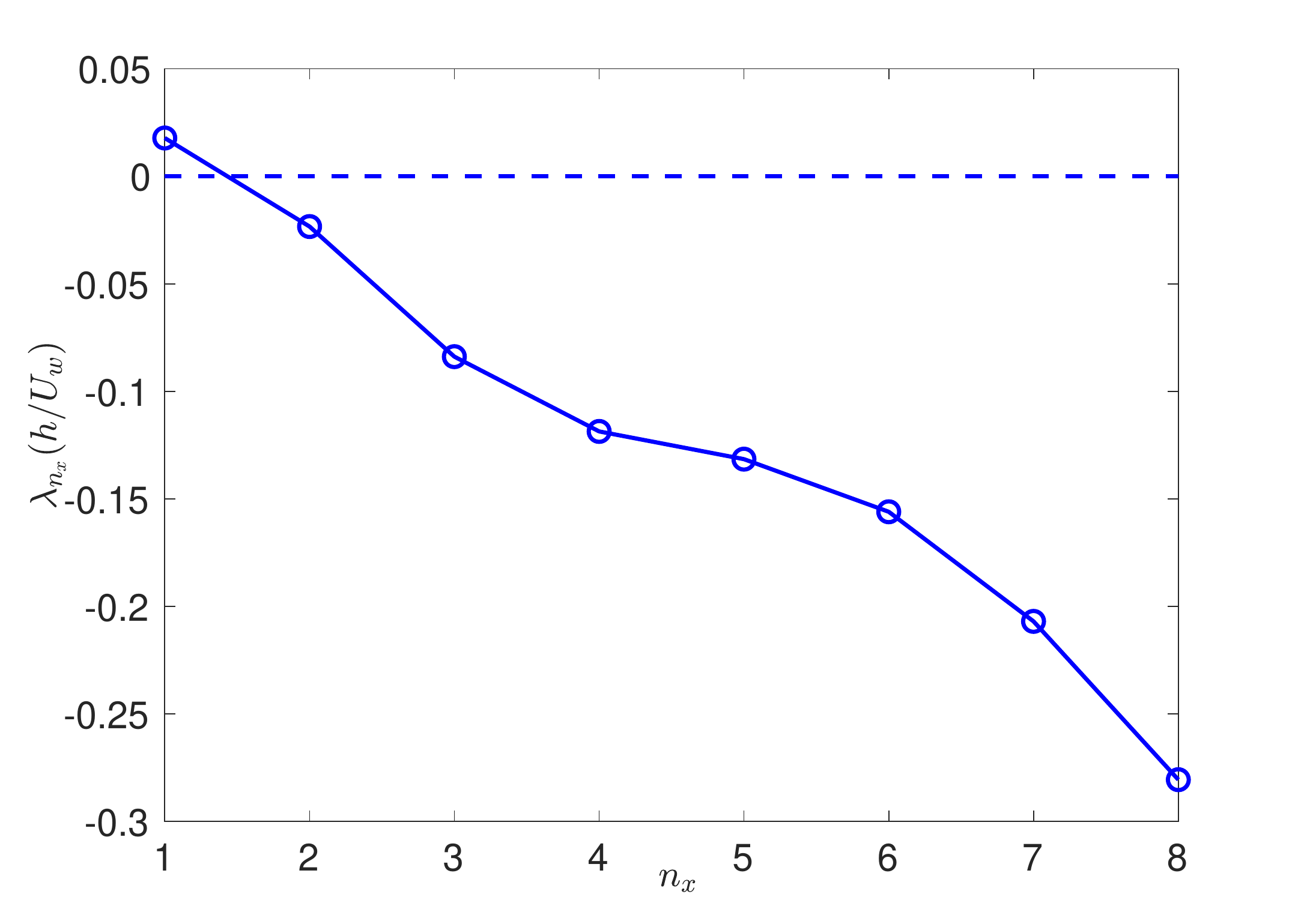}
\vspace{-1em}
\caption{
The top Lyapunov exponent of perturbations  with
channel wavenumbers $n_x=1,\dots, 8$ evolving under the time-dependent 
turbulent mean flow in the DNS, $\U$.
The Lyapunov exponent of all $n_x  \ge 2$ components of $\u'$
is negative.  For comparison, the 
least stable mode of the  streamwise-spanwise-temporal mean of $\U$ has 
decay rate $\sigma=-0.12 ~U_w/h$   at $n_x=1$ and  $n_z = 3$. The Lyapunov 
exponent was nondimensionalized  using advective time units, $h/U_w$. A 
plane Couette channel at $R = 600$ was used.} \label{fig:Lyap_k}
\end{figure*}

We now compare RNL and DNS dynamics in order to gain insight into
the role of perturbation-perturbation nonlinearity $N_4$ in the maintenance and regulation of 
turbulence.
The $N_4$ term  in~\eqref{eq:NSp} 
does not contribute directly to maintaining the perturbation energy because  the  perturbation-perturbation interactions
redistribute energy  internally among the streamwise $k_x \ne 0$ components of the flow and
the term $ \langle \u' \bcdot N_4\rangle_{x,y,z}$ is zero~\footnote{{In our simulations time discretization 
produces a  $\langle \u' \bcdot N_4 \rangle_{x,y,z,t}$ of the order of  $-0.0005 U_w^3/h$ which provides an error estimate for the 
accuracy of our results.}} in the DNS.
From~\eqref{eq:NSp} we obtain that the perturbation energy 
density, $E_p = \langle |\u'|^2 / 2\rangle_{x,y,z}$, evolves according to:
\begin{equation}
 \frac{d E_p}{dt} =    \underbrace{\left \langle \u' \bcdot \left ( -  \U  \bcdot \bnabla  \u ' -
\u ' \bcdot \bnabla  \U   +   \Delta  \u '/R \right ) \right  \rangle_{x,y,z}}_{\dot E_{linear}}~.
%+\langle\u' \bcdot \f \rangle_{x,y,z} ~,%+ \sqrt{\varepsilon} ~{\mathbf{f}}'(x,y,z,t)\ ,
 \label{eq:Ep1}
\end{equation}
just as in RNL turbulence. The  term  $\dot E_{linear}$ comprises   the energy transfer
to the streamwise-varying perturbations by  interaction with the  fluctuating mean 
$\U(y,z,t)$ and the dissipation.
The top Lyapunov exponent of the 
perturbation field $\u'$ associated with the  mean flow taken from
the DNS, as defined in~\eqref{eq:exp}  is also given by
the time-average of the instantaneous perturbation energy growth rates:
\begin{equation}
\lambda_{Lyap} = \left \langle    \frac{1}{2 E_p} \frac{d E_p}{dt} \right \rangle_{t}~.
\label{eq:mle}
\end{equation}
Equation~\eqref{eq:mle} converges asymptotically in $t$ to the top 
Lyapunov exponent for any mean flow and particularly for our analysis for
 the mean flow obtained from the DNS. The 
full spectrum of exponents can also  be obtained using
orthogonalization techniques~\cite{Farrell-Ioannou-1996b}.  

This top Lyapunov exponent should be contrasted with the  exponent obtained
by inserting into~\eqref{eq:Ep1} and obtaining~\eqref{eq:mle}  with 
the $\u'$ taken from DNS.   This $\u'$ is bounded 
because it is the perturbation state vector
and therefore this exponent   is exactly $\lambda_{state}=0$. While only the top Lyapunov 
vector is maintained by the RNL system of our example,
in DNS  a spectrum of Lyapunov 
vectors comprise the $\u'$  of the state and because 
these are orthogonal in energy we can  consider the energetics 
of each of the streamwise Fourier components of $\u'$ separately.
If we decompose the perturbation field into its streamwise components:
\begin{equation}
%\u'  = \sum_{n_x=1}^N ~\u'_{n_x} (y,z,t)  e^{ i k_x x}~,
\u'  = \sum_{n_x=1}^N ~\underbrace{\Re \left ( \av_{n_x} e^{ k_x x} \right )}_{\u'_{n_x}}~,
%
%
%\sum_{n_x=1}^N ~\underbrace{\alpha_{n_x} \cos (k_x x) + \beta_{n_x} \sin (k_x x)}_{\u'_{n_x}} ~,
\end{equation}
with $k_x= 2 \pi n_x /L_x$ and $Re$ denoting the real part, the effective time average growth rate:
\begin{align}
\lambda_{n_x, state} = \Bigg \langle \frac{1}{2 E_{n_x}} \Big\langle \u'_{n_x} &\bcdot \big ( -  \U  \bcdot \bnabla  \u '_{n_x} -
\u '_{n_x} \bcdot \bnabla  \U  \nonumber\\
 &+  \Delta  \u '_{n_x} /R +N_{4, n_x} \big) \Big\rangle_{y,z}  \Bigg \rangle_{t} ~,
\label{eq:Ep2s}
\end{align}
of  each Fourier component $\u'_{n_x}$ of the perturbation field is zero.  In \eqref{eq:Ep2s}
$E_{n_x} = \langle |\u'_{n_x}|^2 / 2\rangle_{y,z}$ is the kinetic energy of the $n_x$ streamwise component and $N_{4, n_x}$ is the $n_x$  
 streamwise perturbation component of  the $N_4$ term  in~\eqref{eq:NSp}.
 In the energetics of DNS  in addition to 
 the rate of the instantaneous energy transfer  to the perturbations from the mean flow:
\begin{equation}
\dot E_{def, n_x} = \left  \langle \u'_{n_x} \bcdot \left ( -  \U  \bcdot \bnabla  \u '_{n_x} -
\u '_{n_x} \bcdot \bnabla  \U    \right ) \right  \rangle_{y,z}~,
\label{eq:edef}
\end{equation}
and the perturbation energy dissipation rate:
\begin{equation}
\dot E_{dissip, n_x} = \frac{1}{R}   \left   \langle \u'_{n_x} \bcdot  \Delta  \u '_{n_x} \right  \rangle_{y,z}~, 
\end{equation} 
which are the only terms  in~\eqref{eq:Ep1} involved in the determination of the Lyapunov exponent, the additional term
\begin{equation}
\dot E_{nonlin,n_x} =  \left   \langle \u'_{n_x} \bcdot  N_{4, n_x} \right  \rangle_{y,z}~, 
\end{equation} 
giving the net energy transfer rate at each instant to the other nonzero streamwise
components  appears in the DNS equations.

For convenience we define the linear operator $A$ so that
$$\u'  \bcdot \left ( -  \U  \bcdot \bnabla  \u '-\u ' \bcdot \bnabla  \U    \right ) \equiv \u' \bcdot ( A +A^\dagger ) \u' /2\ ,$$
where $A^\dagger$ is the adjoint operator in the energy inner product. The eigenvalues of the linear operator
$(A+A^\dagger)/2$  order in the orthonormal basis of the  eigenfunctions of this operator
the rate of transfer of energy from the instantaneous streamwise mean flow to the perturbations.

\begin{figure*}
\centering\includegraphics[width=30pc]{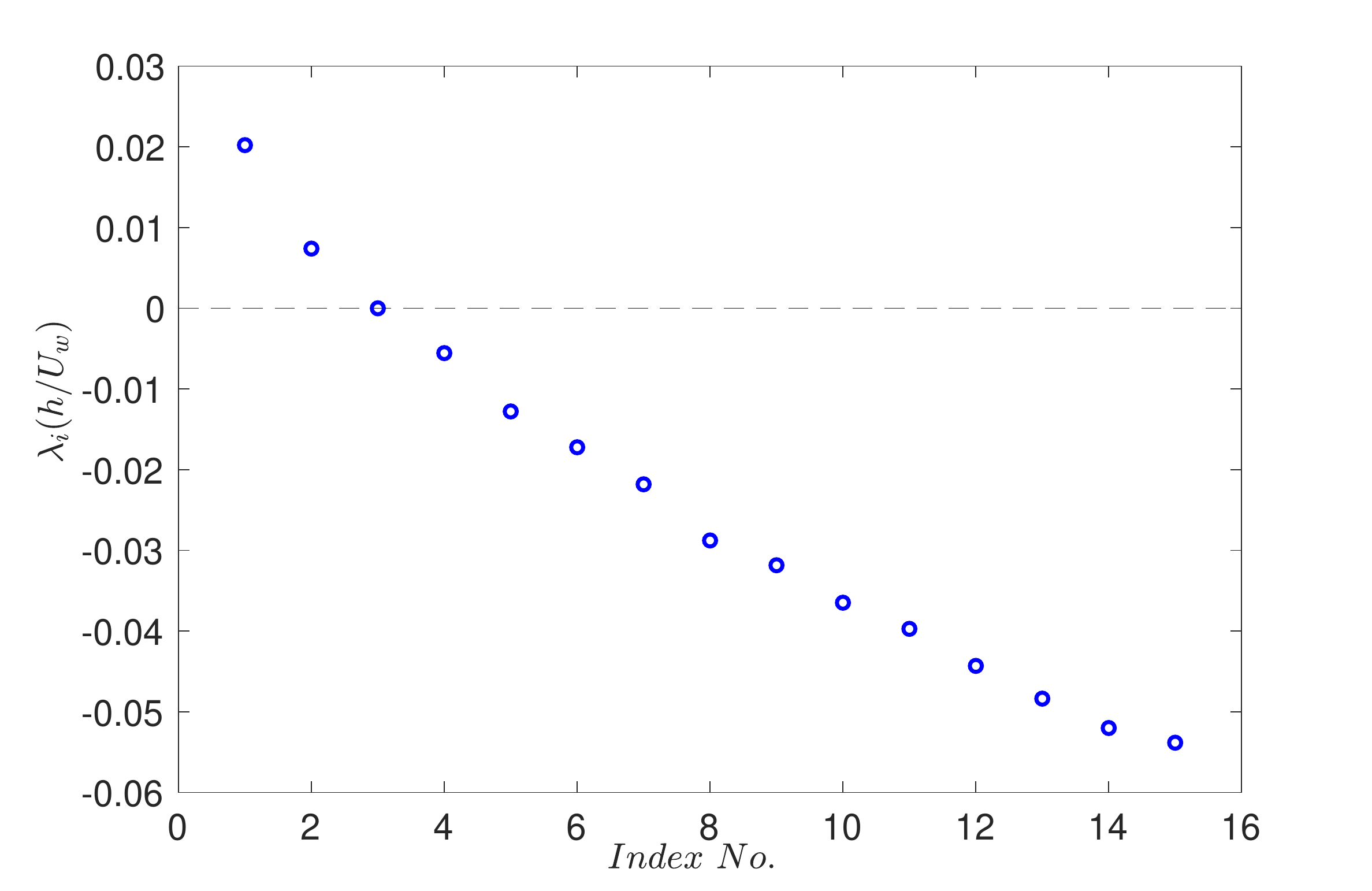}
\vspace{-1em}
\caption{The first 15 Lyapunov exponents of the DNS mean flow for $n_x=1$. } \label{fig:lyaps1}
\end{figure*}

The dynamical significance  of  $N_4$ in sustaining the turbulent state is revealed  by comparing
the perturbation energetics  under the influence of  the DNS mean flow $\U$ with and without the term $N_4$.
This can be achieved  by calculating  the Lyapunov exponents $\lambda_{Lyap}$ of  
the $\U$ obtained from a DNS and the associated Lyapunov vectors 
together with  the   contributions of each of these Lyapunov vectors to the energy transfer rates
 $\dot{E}_{def,n_x}$,   $\dot{E}_{dissip,n_x}$, $\dot{E}_{nonlin,n_x}$  and  comparing these rates 
with and without  the term $N_4$.  Although the $N_4$ term   
is energetically neutral   it
may have a profound impact on the energetics by modifying  the perturbations to
extract more or less energy from the
mean flow. Evidence that  the term $N_4$ is not fundamental to sustaining the turbulence   
 but instead the parametric mechanism of RNL is fundamentally responsible for maintaining DNS turbulence 
 would be provided by the following four conditions: 
 {(\emph i)}
 the top Lyapunov exponent, $\lambda_{n_x}$, is  associated with the same streamwise components
$n_x$ of the turbulent field in RNL and DNS and is  neutral after accounting for the transfer of energy
to the other streamwise perturbation components, which would indicate
that the turbulent state is regulated to 
neutralize the top Lyapunov vector growth rate (maximum Lyapunov  exponent)
 coincident with the (necessary) neutralization of the state vector, 
  ($\emph ii$) the transfer of energy from the mean flow by the top Lyapunov vector
   $\dot{E}_{def,n_x}$  should exceed that by the state 
  vector indicating that $N_4$ has disrupted the Lyapunov vector 
  making it less effective at transferring energy from the mean 
  flow,  ($\emph iii$) the Lyapunov vectors
span the energy and the energetics of the DNS perturbation field in a 
convincingly efficient manner, 
most tellingly if they span it in the order of their growth rate,
and ($\emph  iv$) in addition to supporting the perturbation energy 
and energetics the Lyapunov vectors  support the roll circulation 
maintaining the coherent roll/streak structure.

Satisfying these conditions
would   strongly support the conclusion that the DNS turbulence 
is being maintained  primarily through the parametric perturbation growth 
process associated with the temporal variation of $\U$ that supports turbulence in RNL, 
without substantial contribution from the $N_4$ nonlinearity.
The 
alternative is that the $N_4$ term contributes at leading order to
the energetics which would imply centrality in the dynamics of turbulence for
the alternative role for $N_4$,  which is to   replenish the subset of perturbations lying in 
the directions of growth.
%as   in a number of toy models  of turbulence dynamics
%\citep{Trefethen-etal-1993,Gebhardt-Grossmann-1994,Baggett-Trefethen-1997,Grossmann-2000}.  
This distinction in mechanism can be clarified by observing that, if instead of 
making the dynamically crucial choice of the streamwise average 
as the mean in constructing the RNL system and the mean flow were instead
chosen to be
the time-independent streamwise-spanwise-temporal mean,  which in a boundary layer 
flow is the stable Reynolds-Tiederman profile~\cite{Reynolds-Tiederman-1967},
the $N_4$ nonlinearity must assume this role if turbulence is to be 
sustained.  This follows because the alternative 
 parametric mechanism would not be available. Turbulence could in principle
%in sustaining the turbulent state,  and  statistically stationary 
%turbulence 
be sustained by this mechanism if $N_4$ were  
%for this streamwise-spanwise-temporal mean-perturbation decomposition
sufficiently effective in scattering  perturbations back into the 
directions of non-normal growth. %However, it is known that this is not the case~\citep{Jimenez-Pinelli-1999}.
% as is commonly hypothesized  in toy models
%\citep{Trefethen-etal-1993,Gebhardt-Grossmann-1994,Baggett-Trefethen-1997,Grossmann-2000}.  
However, the experiments of Jimenez \& Pinelli \cite{Jimenez-Pinelli-1999} show 
that this is not the case.  They demonstrate
that removing the  streak component in a DNS of a channel flow laminarizes the flow.
Although the mean flow in their DNS remains highly
non-normal the  mechanism of nonlinear scattering back 
into the growing subspace of the non-normal operator can not  maintain a turbulent state in the absence parametric 
mechanism made available by the 
specific structure of the fluctuating streak.
%This shows that
%certain nonlinear interactions  that were removed with the suppression of the 
%streak  are crucial for the maintenance of the turbulence state   \cite{Waleffe-1997} and that 
%the nonlinearity viewed generically is not a random  mixer. The second order
%closure SSD with the averaging operator defined  as the streamwise average presents  a minimal representation
%of the turbulent state that includes the necessary ingredients of nonlinearity.}
%

\begin{figure*}
\centering\includegraphics[width=30pc]{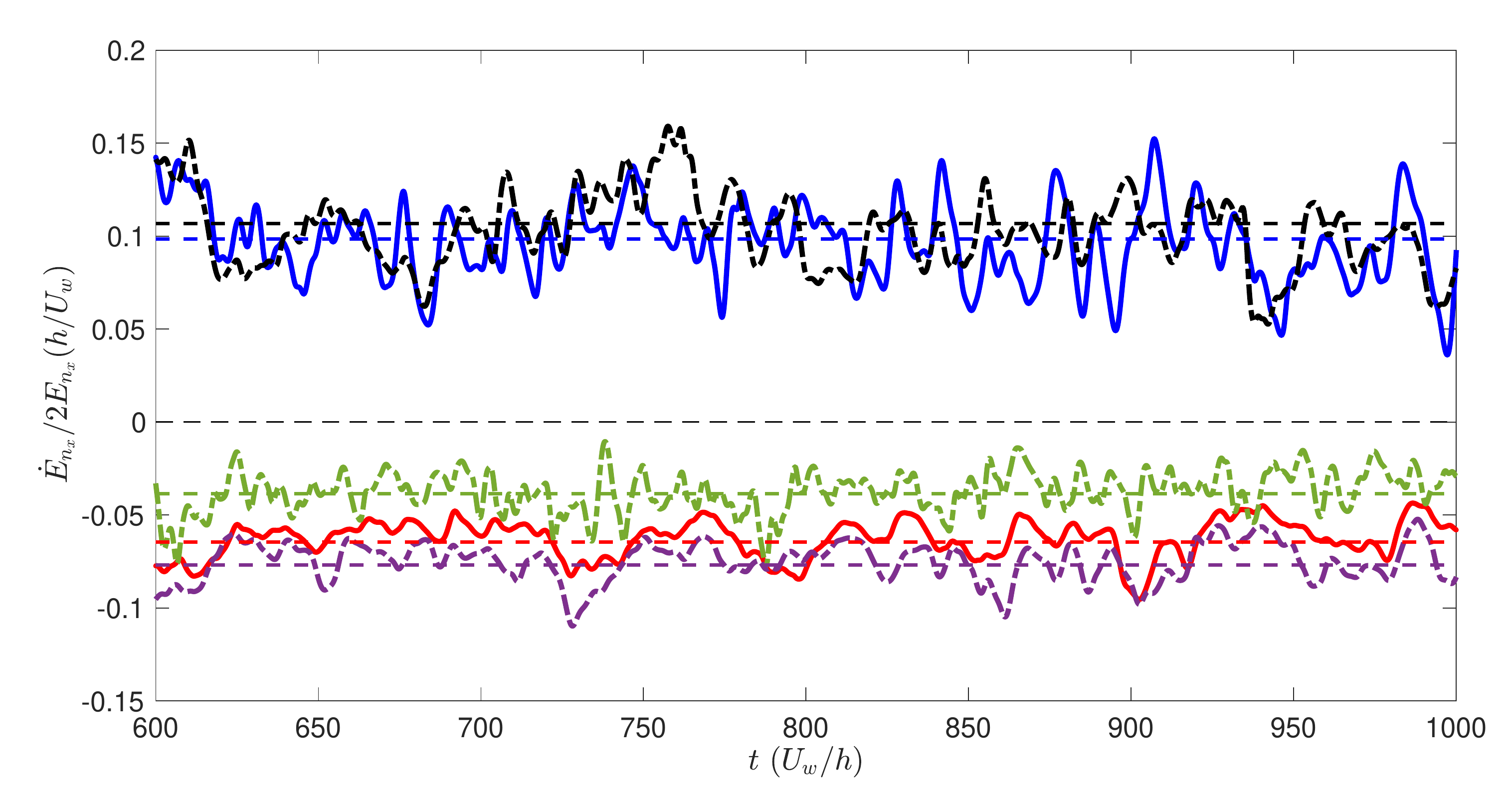}
\vspace{-1em}
\caption{Contribution to the instantaneous energy  growth rate of the $n_x=1$ perturbation component in DNS: extraction 
from the fluctuating $n_x=0$ mean component $\dot E_{lin, n_x}/(2 E_{n_x})$ (blue, solid); 
loss to dissipation $\dot E_{dissip, n_x}/(2 E_{n_x})$ (red, solid); transfer to the other $n_x >1$ streamwise 
components $\dot E_{nonlin,n_x}/(2 E_{n_x})$ (green).
The mean values of these rates are indicated with the dashed lines with the corresponding color. These average rates sum to $\lambda_{state}=0$.
The corresponding rates for the first
Lyapunov vector
are shown in black and purple dash-dotted lines (there is no energy transfer to the other components as $N_4$ is absent in this calculation). 
These rates sum to the
Lyapunov exponent $\lambda_{Lyap} =0.02 U_w/h$, which is as expected positive and comparable  to
the transfer rate from the state vector,  $\dot E_{nonlin,n_x}/(2 E_{n_x})=0.04 U_w/h$. }
%The brown curve  gives the  rate of energy extraction of the mean by the
%projection of the NL state on the first 30 fastest growing Lyapunov vectors, $\dot E_{30,n_x}/(2 E_{30,n_x})$.
%The green curve indicates the rate of transfer from $n_x=1$ to the other streamwise components.
%This figure shows that the nonlinear interactions perturbation - perturbation nonlinearity, 
%N_4, disrupts the parametric energetic interaction 
%of the perturbations with the mean rather than configuring the perturbations so as to extract  
%more energy from the mean flow.}
% and that
%the energetically active subspace of the perturbations is effectively spanned  by the first 15 Lyapunov vector pairs.}
 \label{fig:comp1}
\end{figure*}

% \clearpage

\section{The Lyapunov exponent of the mean flow in Couette turbulence  at R=600}

% \begin{figure*}
%\centering\includegraphics[width=30pc]{Figures/Lyap_DNS-eps-converted-to.pdf}
%\vspace{-1em}
%\caption{Converging timeseries of the $\lambda$ exponent corresponding to the state and Lyapunov instantenteous growth rates} \label{fig:L1_state}
%\end{figure*}

\begin{center}
\begin{table}
%\caption{}
\caption{\label{table:geometry}The channel is periodic in the streamwise, $x$, and spanwise, $z$, direction
and at the channel walls $y=\pm h$ the velocity is $\u = (\pm U_w, 0, 0)$. The channel length is  $L_x$ and $L_z$  in the streamwise and spanwise directions respectively.
The number of streamwise and spanwise Fourier components is  $N_x$ and $N_z$
after dealiasing  in the streamwise and spanwise direction by the $2/3$ rule,
and we use  $N_y$ grid points in the wall-normal direction.
%$\Ret$is the Reynolds number of the simulation based on the friction velocity.
$R = U_w h / \nu$ is the Reynolds number of the simulation, with $\nu$ the kinematic viscosity.
}%(Add $Re_{bulk}$)}
%and $[L_x^+$,$L_z^+]$ is the channel size in wall units.}
%%\footnotesize\rm
\centering\vspace{.8em}
\begin{tabular}{@{}*{5}{c}}
\break
 Parameter  & $[L_x,L_z]/h$ &$N_x\times N_z\times N_y$& $R$ \\
 NS600   & $[1.75\pi\;,\;1.2\pi]$&$17\times 17\times 35$&600   \\
% RNL100  & $[4\pi\;,\;\pi]$&$ 3\times 63\times 97$&$100.98$&1950& $Re_\tau$ &$100.59$\\
\end{tabular}
\end{table}
\end{center}

\vskip-0.2in
Consider a Couette turbulence simulation at $R=600$ in a  periodic channel with parameters given in Table \ref{table:geometry}.
This is a larger channel than the minimal Couette flow  channel studied by 
Hamilton, Kim \& Waleffe~\citep{Hamilton-etal-1995} at $R=400$.
RNL turbulence with these parameters at $R=600$
was systematically examined recently
\citep{Farrell-Ioannou-2017-sync}.
%The perturbation state in RNL turbulence is supported by the set of Lyapunov vectors that have been regulated by feedback between the streamwise mean and perturbations to have zero Lyapunov exponent~\citep{Farrell-Ioannou-2012,Thomas-etal-2015,Farrell-etal-2016-VLSM}.  
%At the Reynolds  number $R=600$
%used in this study there is only one such Lyapunov vector (with twofold degeneracy) so the entire support of the turbulent state is by this vector.  However, under either a stochastic parameterization for N_4 or, as in the present study, inclusion of the N_4 term in the dynamics, the perturbation energy becomes supported by a larger set of structures due to nonlinear energy scattering~\citep{Farrell-Ioannou-2017-sync}.  We are interested to determine whether the primary support of
%both the energy of the perturbations and the energy transfer from the mean to the perturbations 
%rcontinues to be the first few lyapunov vectors, as it is in the case of RNL turbulence, 
%%continues to be by the Lyapunov vectors 
%which would be expected if the parametric mechanism was the primary nonlinear mechanism supporting the NL turbulence as it is for the support of the RNL turbulence.
%

We first calculate the
Lyapunov exponent $\lambda_{Lyap}$
of the DNS streamwise mean flow by estimating~\eqref{eq:mle} from 
a long integration  of~\eqref{eq:RNSp1} with the mean flow $\U$  obtained from
a turbulent DNS.
The initial state $\u'$ is inconsequential
because,  with  measure zero exception,  any random initial condition converges 
in this system with exponential accuracy
to the Lyapunov vector 
associated with the largest  Lyapunov exponent. The full spectrum of Lyapunov exponents and vectors
can be obtained by an orthogonalization procedure.
% with the inner product of two flow fields, $\u_1'$ and $\u_2'$, defined as the volume integral of the dot product of the velocities:
%$\langle \u_1' \bcdot \u_2' \rangle_{x,y,z}$. 
For a  discussion of the calculation and properties of
Lyapunov exponents and the  associated Lyapunov vectors refer to Refs.~\citep{Farrell-Ioannou-1996b,Farrell-Ioannou-1999, Farrell-Ioannou-2017-sync, Wolfe-Samelson-2007,Cvitanovic-etal-2016}. Because of the streamwise
independence of $\U$,  the different streamwise Fourier components of $\u'$ in this Lyapunov exponent calculation, in which the
 $N_4$ term is absent, evolve
independently and the Lyapunov vector associated with a given Lyapunov exponent
has streamwise structure confined to a single  streamwise  wavenumber $k_x = 2 \pi n_x  h/ L_x$, 
corresponding to the $n_x$ streamwise Fourier component.

The top Lyapunov exponent  at each $n_x$ is shown in figure~\ref{fig:Lyap_k}. This
plot  reveals 
 that the time dependent streamwise mean flow 
$\U$  is  asymptotically stable to all perturbations with $n_x>1$ with only  the $n_x=1$ streamwise component  supporting
a positive Lyapunov exponent of
$\lambda_{Lyap}\approx 0.02 U_w/h$.  
  %or $\lambda_{Lyap} \approx 0.4 u_\tau / h$.  
  Recall that in RNL the top Lyapunov exponent also has wavenumber  
  $n_x=1$ and is exactly 
  zero  consistent with  mean $\U$ being adjusted by  feedback 
  through the Reynolds stress term $N_2$ 
to   exact  neutrality.  The top Lyapunov exponent  obtained using  the $\U$ of  DNS
is  positive consistent with the requirement to account for 
the  energy  exported to other perturbations.
%The degree of  positivity  of $\lambda_{Lyap}$ is necessarily adjusted by feedback 
%through $N_2$ to account for  this loss. With this consideration  the DNS mean flow $\U$
%can be verified to be neutral. 
Figure~\ref{fig:lyaps1} shows that the DNS mean flow being without 
energy loss to scattering by the N4 term actually supports
two positive  Lyapunov exponents with $n_x=1$.

\begin{figure*}
\centering\includegraphics[width=30pc]{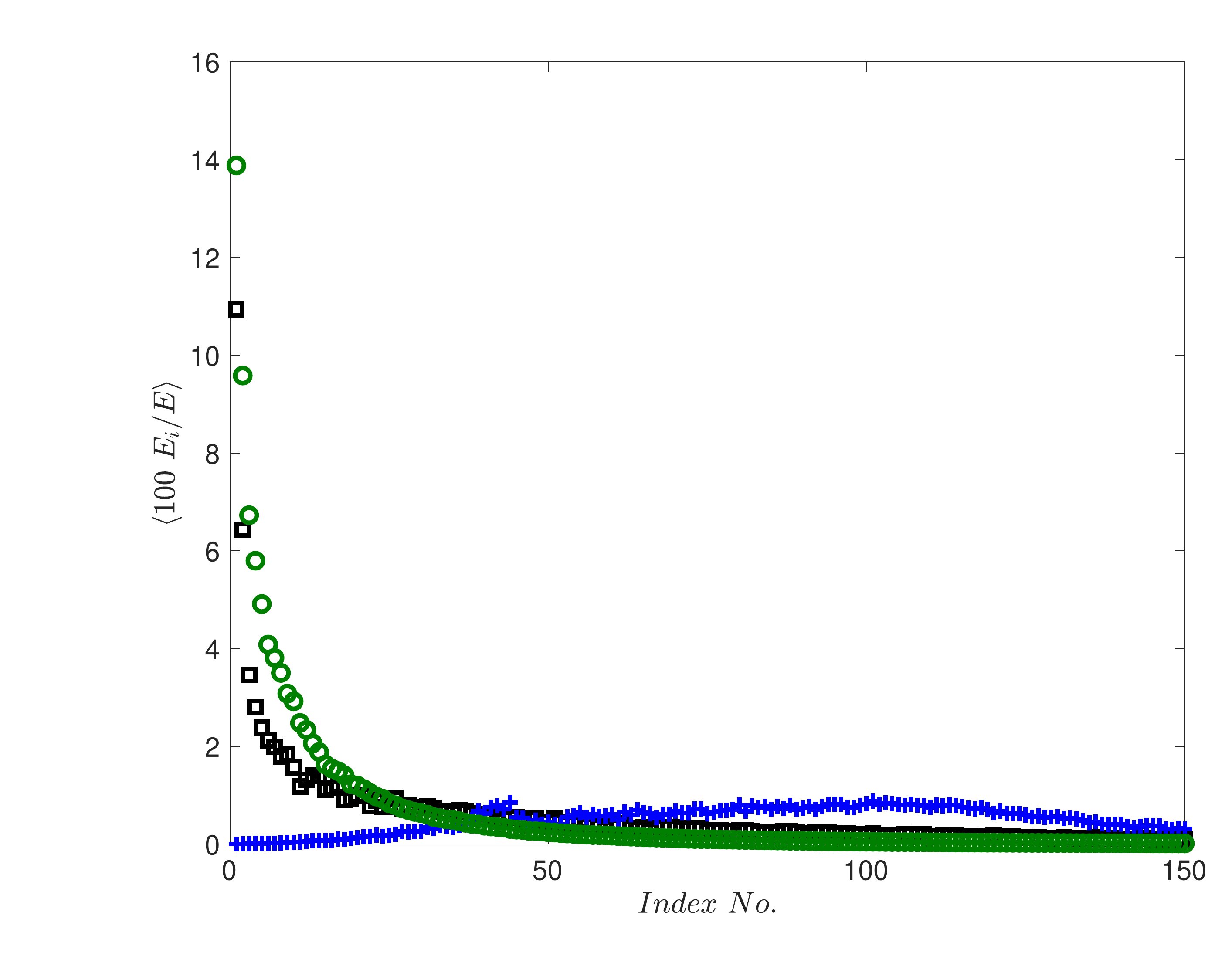}
\vspace{-1em}
\caption{Average energy percentage of the $n_x=1$ flow accounted for by each  Lyapunov vector (black squares). 
Average energy percentage of the $n_x=1$ flow accounted for by the eigenvectors of  $ A+A^{\dagger}$ 
ordered in descending order of their eigenvalue (blue crosses). 
%Average energy percentage of the $n_x=1$ flow accounted for by the eigenvectors of  $\langle A+A^{\dagger} \rangle_t$ 
%ordered in descending order of their eigenvalue (red diamonds). 
$A$ is the  operator in 
\eqref{eq:RNSp1} governing the linear evolution of the perturbations about $\U$.
%With blue crosse is the average energy percentage of the $n_x=1$ flow accounted for by the eigenvectors of  
%of  $A+A^{\dagger}$ 
The eigenvectors of $ A+A^{\dagger}$ 
are the orthogonal directions of stationary  instantaneous perturbation energy growth rate, with this 
 growth rate 
given by the corresponding eigenvalue. The perturbation component of the turbulent flow 
is adjusted to have a small projection on the first eigenvectors of $A+A^{\dagger}$
 associated with large instantaneous energy growth rates.  
 Also shown is the energy percentage accounted for by the PODs of the $n_x=1$ component
 of the NL perturbation state (green circles).
 } \label{fig:proj}
\end{figure*}

%\begin{figure*}
%\centering\includegraphics[width=28pc]{cumulative_simple_1-eps-converted-to.pdf}
%\vspace{-1em}
%\caption{ Average fraction of the energy  of the $n_x=1$ flow accounted for by the first $N$ pairs of Lyapunov vectors (black),
%%by the first $N$  eigenvectors of  $\langle A+A^{\dagger} \rangle_t$ 
%%(red), 
%by the first  $N$ pairs of  eigenvectors of  
%of  $A+A^{\dagger}$ (blue)  ordered in descending order of their eigenvalue, 
%and by the first $N$ pairs of POD's of the $n_x=1$ component of the NL state (green). 
%} \label{fig:cum}
%\end{figure*}

Contributions to the Lyapunov exponent 
from  mean flow energy transfer and from dissipation
are plotted as a function of time in figure~\ref{fig:comp1}.  The growth rate associated with energy transfer
from the fluctuating streamwise  mean is on average $0.11 U_w/h$, while the dissipation rate is on average
$0.09 U_w/h$ resulting in the positive  Lyapunov exponent $\lambda_{Lyap} = 0.02 U_w/h$.
These transfers occur
when perturbations evolve under the dynamics of the fluctuating streamwise mean flow $\U$ of the DNS but 
in the absence
of two effects: ($\emph i$) 
disturbance to the perturbation structure by the perturbation-perturbation nonlinearity $N_4$ 
and ($\emph ii$) transfer of energy to other perturbations by $N_4$.  
This result demonstrates that the parametric growth mechanism is able to 
maintain the perturbation turbulence component against dissipation 
with additional energy extraction to account for transfer to the other scales.  

\begin{figure*}
\centering\includegraphics[width=30pc]{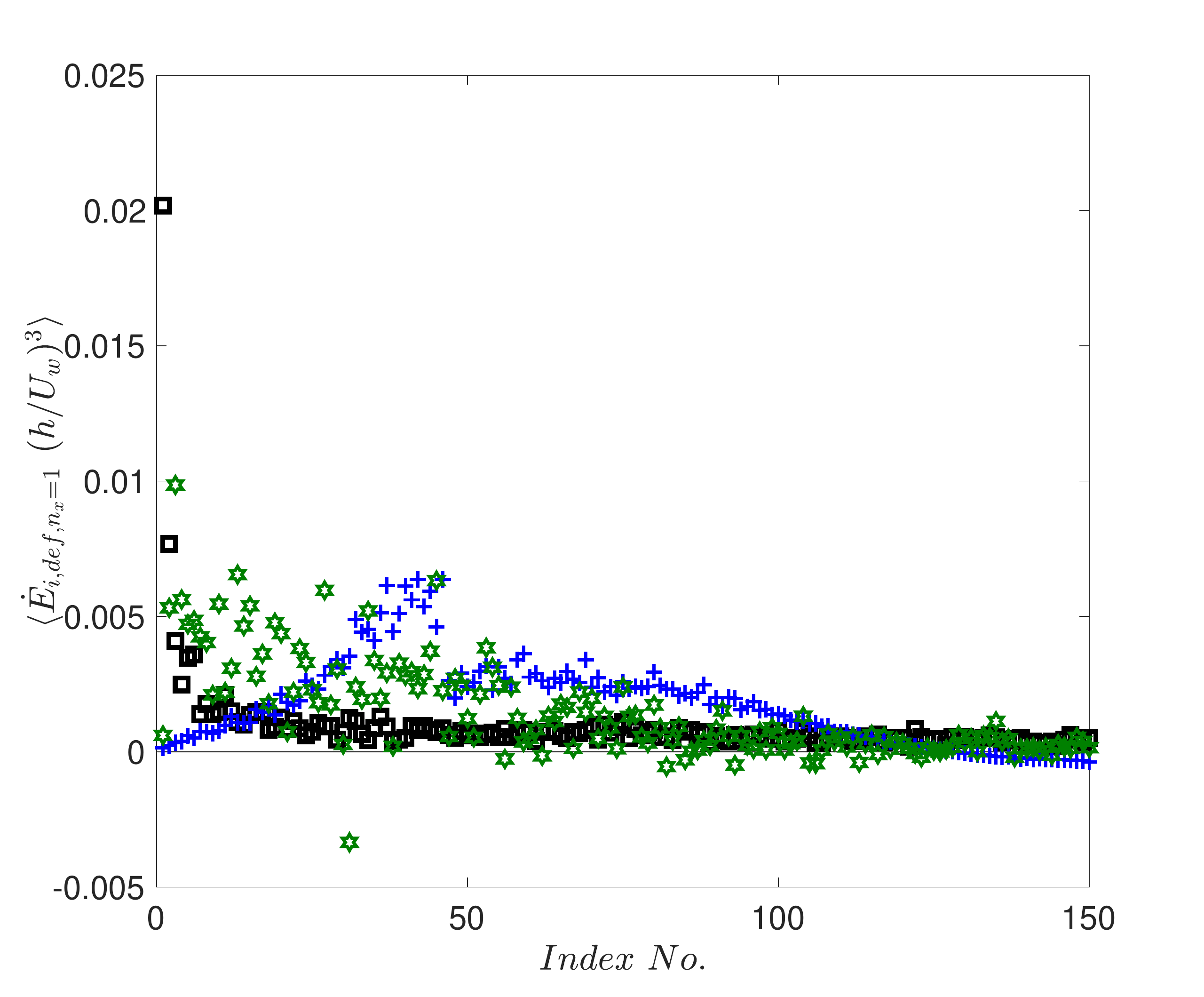}
\vspace{-1em}
\caption{  Contribution of the  Lyapunov vectors (black squares), the PODs (green stars) and  the eigenvectors
of $A+A^\dagger$ (blue crosses) to the energy transfer rate, $\dot E_{lin}$,   from the mean flow. 
 } \label{fig:edot}
\end{figure*}

%\begin{figure*}
%\centering\includegraphics[width=30pc]{a_time-eps-converted-to.pdf}
%\vspace{-1em}
%\caption{  \color{blue}Time series of the square of the projection coefficients $a^2(t)$ of the NL perturbation state on the first three Lyapunov vectors.
%The state projects predominantly on the first two Lyapunov vectors. However, because the top Lyapunov vectors extract 
%energy from the mean flow at nearly the same rate on average, the NL perturbation state extracts  energy at the same rate on average
%with the first Lyapunov vector as shown in figure~\ref{fig:comp1}. } \label{fig:atime}
%\end{figure*}

We now contrast the energetics of the Lyapunov vectors on the  DNS mean flow just shown
with the corresponding energetics of the
 $n_x=1$ Fourier component of the
state vector obtained from the  DNS itself  in order to  determine
whether the $N_4$ term has the effect of influencing the perturbations to have a more or less
favorable configuration for extracting energy from the mean flow.
These results are also shown in figure~\ref{fig:comp1} from which it can be seen that
although the DNS turbulent state vector episodically exceeds its associated Lyapunov vector 
in rate of energy transfer from the mean flow this transfer rate
 with the influence of the $N_4$ term included
is slightly less on average than that  achieved by the first Lyapunov vector in the absence of the influence of $N_4$:  
energy transfer rate to the
DNS state vector produces  growth rate $0.099 U_w/h$  compared to  $0.107 U_w/h$
for the  Lyapunov vector on the DNS mean flow.
This demonstrates that the nonlinear term $N_4$ does not configure the perturbations
 to transfer more energy from
the highly non-normal  streamwise mean flow $\U$ on average.   However, despite 
the fact that the energy transferred from the streamwise mean flow by the DNS
perturbation state and  by the  first Lyapunov vector are nearly equal when averaged over time, the correlation 
coefficient of the transfer rate time series,  
shown in figure~\ref{fig:comp1}, is low ($0.26$) suggesting
 that the $N_4$ term has disrupted the first Lyapunov vector 
 and spread its energy to other Lyapunov vectors.
The fact that this disruption does not substantially alter
the time-mean energy transfer from the streamwise mean flow 
suggests that the time mean energetics resulting from projection on the Lyapunov vectors of 
$\U$  is not substantially altered by $N_4$ while the projection at an instant in time is.
%(recall that in RNL $N_4$ is absent and this projection is onto the single top Lyapunov vector).  
 Note also that in the energetics
 of the $n_x=1$ perturbation component  in DNS  there is a term not present in the corresponding Lyapunov vector:
 the energy interchanged with  the remaining $n_x\ne 0$ components, which is also shown in figure~\ref{fig:comp1}.
 The $n_x=1$ perturbation component of the DNS exports energy to the other streamwise components of the flow
 and this  transfer contributes $0.04 U_w/h$ at this wavenumber to the decay rate.
 This additional decay is just sufficient to reduce
 the  mean growth rate of the DNS
 to the required value $\lambda_{state}=0$. 
 
 We conclude that the Lyapunov exponent of the fluctuating streamwise mean flow 
 $\U$  in DNS turbulence  has been adjusted to near  neutrality and with energetics consistent 
 with the parametric growth mechanism
 fully accounting for the maintenance of the 
 perturbation component of the turbulent state. The perturbation-perturbation nonlinearity, $N_4$,
 does not configure the perturbations to extract more energy from the streamwise mean flow  than 
 in the absence of this term, implying
 that  $N_4$ acts as
a negative influence  on the perturbation growth process. 
This is opposite to the mechanism 
in toy models of turbulence in which nonlinearity systematically 
configures perturbations to be more effective at exploiting the non-normality of the mean flow
(cf. \cite{Trefethen-etal-1993}).
 The fact that the mean DNS flow has been adjusted to near neutrality of its first Lyapunov 
 vector  suggests this structure is 
 controlling the parametric instability of the mean state and therefore that the  first Lyapunov vector should be a dominant component 
of the perturbation state in the DNS. However,  differences between the time series of the energy transfer 
rate by the DNS state vector and by the top Lyapunov vector of the associated mean state 
suggests that other (decaying) Lyapunov vectors have been excited by $N_4$.
This will be examined in the next section.

%\section{Perturbation state decomposition in a Lyapunov vector basis}
\section{Analysis of perturbation energetics by projection onto  the Lyapunov vector basis}
\vskip0.2in

Despite the correspondence between the mean energetics of the DNS 
perturbation state and the mean energetics of the top 
Lyapunov vector calculated using the associated fluctuating streamwise mean 
flow it remains to explain why time series of perturbation growth rate for these shown in figure~\ref{fig:comp1}
%and snapshots of the perturbation state and the Lyapunov vectors shown in figure~\ref{fig:snap}
reveal considerable differences. 
This suggests further analysis to clarify the relation between  the perturbation state and the Lyapunov vectors. 
The orthogonality property imposed on the Lyapunov vectors makes them an attractive basis for analyzing the relation between 
perturbation structure and energetics.  Expanding the $n_x=1$ DNS perturbation state 
$\hat \u'$ in the basis of the orthonormal in energy $n_x=1$ Lyapunov vectors, $\u_i'$:  
\begin{equation}
\hat \u' (t) = \sum_i a_i(t) \u _{i}'(t),
\label{eq:expand}
\end{equation}
with  projection  coefficient:
\begin{equation}
a_i(t)=\left \langle \u _{i}'(t)\bcdot \hat \u' (t)\right \rangle_{x,y,z},
\end{equation}
we obtain that  the contribution to the perturbation energy of Lyapunov vector $\u_{i}'$ 
is $E_i=a^2_i(t)/2$.
%Due to the existence of degenerate pairs
Projection of the energy of the $n_x=1$ component of the perturbation state on the first 150 Lyapunov vectors is shown in figure~\ref{fig:proj}. 
The percentage of energy accounted for by projection on  the most unstable
%of the $n_x=1$
%state contained in the first 
Lyapunov vector
is $11\%$, significantly larger than the energy in each of the remaining Lyapunov  vectors. 
Adding the second unstable
Lyapunov vector raises this value to $17.4\%$ and the first 100
$n_x=1$ Lyapunov vectors account for $82 \%$ of the energy of the $n_x=1$ component 
of the perturbation state. In order to understand the significance of the Lyapunov vectors as a basis for representing the 
DNS  perturbation state we have determined the orthonormal structures  of the proper orthogonal decomposition  (PODs) of
the $n_x=1$ component of the DNS  with the methods discussed in Ref.~\citep{Nikolaidis-etal-Madrid-2016}.
Comparison of the perturbation energy projected on the Lyapunov and the canonical POD basis in Figure~\ref{fig:proj} % and Figure~\ref{fig:cum}
demonstrates that the Lyapunov vectors provide a good representation of the DNS perturbation state. 
 We note that the energy of the perturbation state is partitioned into the Lyapunov vectors
 in the order of their Lyapunov exponent,
while the energy accounted for by the  POD basis
necessarily decrease monotonically  with the order  
of the POD this is not required of the Lyapunov vectors and therefore this monotonic 
decrease provides evidence that the Lyapunov vectors are active agents in the perturbation energetics.
On the other hand,  note that the Lyapunov vectors are not constrained
by optimality of the POD basis to be an inferior basis for spanning the energy, because the 
Lyapunov vectors are time dependent and could theoretically span all the perturbation
energy, as indeed is the case in RNL  for which the entire energy and energetics is accounted for
 by the first Lyapunov vector.

In figure~\ref{fig:proj} we also show the average projection of the DNS state on the
eigenvectors  of the operator $ A+A^{\dagger} $
 ordered in descending order of their eigenvalues. $A $ is the linear operator in 
\eqref{eq:RNSp1} governing the evolution of the perturbations, $\u'$, about $\U$. The eigenvectors of $A+A^{\dagger}$
form an orthonormal set of perturbation structures ordered decreasing in instantaneous energy growth rate 
in  the flow, $\U$.  Typically in turbulent flows both the Lyapunov vectors and the 
perturbation state have
 small projection on the first eigenvectors of $A+A^{\dagger}$, which are the 
 structures producing greatest  instantaneous energy growth rates. 
 The turbulent mean flow $\U$ is such that
 perturbations that lead to large instantaneous growth rate have large 
wavenumber and are located in episodically occurring regions of high deformation.
The top Lyapunov and state vectors are instead concentrated at larger scale
with relatively small  instantaneous growth rate. Small projection of the perturbation state on the directions of maximum instantaneous growth rate was previously seen in RNL simulations at $R=600$
\citep{Farrell-Ioannou-2017-sync}. What is remarkable and indicative of the fundamental 
role of the Lyapunov vectors in the dynamics of DNS is the ordering 
of the perturbation energy in the Lyapunov vectors.
Despite the dynamic importance of the basis of the eigenvectors of  $A+A^\dagger$
comparable ordering does not occur for this basis.

%Also shown are projections
%of the state on the directions of the eigenvectors of the instantaneous
%operator $A+A^{\dagger}$. 

 In RNL simulations at $R=600$ the perturbation turbulent  state is
 entirely supported by the top Lyapunov vector and the energetics of the perturbation state
 consequently are the energetics of this single Lyapunov vector. The $N_4$ nonlinearity distributes  
 the perturbation energy over a subspace spanned primarily by the leading
  Lyapunov vectors, as shown in figure~\ref{fig:proj}.
 We can determine the distribution of the first $N$ Lyapunov vectors ordered in contribution to the perturbation state energy growth 
 rate, $\dot E_{i, def}$ 
 by calculating
  \begin{equation}
\sum_{i=1}^N \dot E_{i , def} \equiv \left \langle \u'_< \bcdot \left ( -  \U  \bcdot \bnabla  \u '_< -
\u '_< \bcdot \bnabla  \U    \right ) \right  \rangle_{x,y,z,t}~,\label{eq:edefn}
\end{equation}  
where 
\begin{equation}
\u'_< (t) = \Re \left ( \sum_{\alpha=1}^N a_\alpha(t) \u' _{\alpha}(t) e^{ 2 \pi i x /L_x} \right ),
\end{equation}
 is the projection of the $n_x=1$ perturbation state, given in~\eqref{eq:expand}, on the first $N$ Lyapunov vectors.
 From this calculation, we can  obtain the incremental contribution to the perturbation energy growth, $\dot E_{i, def}$, of each Lyapunov vector. 
 We can similarly determine the contribution of each of the eigenvectors of $A+A^{\dagger}$ and of the PODs to the 
 energetics of the perturbation state.  
%  The concentration of the energetics around the 40th mode of 
% $A+A^{\dagger}$ indicates the regulation of the mean flow by the turbulence to excite the  structures
% that on the mean have zero instantaneous growth. The  same picture arose in RNL~\cite{Farrell-Ioannou-2017-sync}.
The results, shown in Fig.  \ref{fig:edot}, reveal that the  Lyapunov vectors 
provide the primary support for
the perturbation energetics and their energetic contribution follows the Lyapunov vector growth rate ordering.
If the $N_4$ term were dominant in determining the structures supporting the perturbation state the energetics of the
turbulent state in the DNS
would not be expected to so closely reflect the 
asymptotic structures of the Lyapunov vectors.  
Also note that the first 70 PODs, which contain $95 \%$ of the perturbation energy,
are responsible for most of the energetic transfers, but their contribution is not ordered as in the case of the 
Lyapunov vectors, actually it is almost white.
Note also that the contribution of the first 45 eigenfunctions of $A+A^\dagger$ is  in reverse order of their
instantaneous growth rate.

% \begin{figure}
%\centering
%\includegraphics[width=\columnwidth]{Ox_Oxdot_Ox2_half_obs}
%\vspace{-2em}
%\caption{Time evolution of the contribution of the torques arising from the Reynolds stresses produced by  $\u'_<$  to  maintenance of $\Omega_x^2$, which largely consists of the rolls (blue). Their mean contribution to roll maintenamce is $0.056~ h/U_w$. 
%The torques from $\u'_>$ (red) make a mean contribution to the rolls of $0.2 h/U_w$ in this measure.
%Their time mean  is indicated with  dashed lines.
%The contribution of the top 4 LVs to the roll maintenance is underestimated  in this measure because this field 
%is highly structured. } 
%\label{fig:Ox}
%\end{figure}

 \begin{figure}
\centering
\includegraphics[width=\columnwidth]{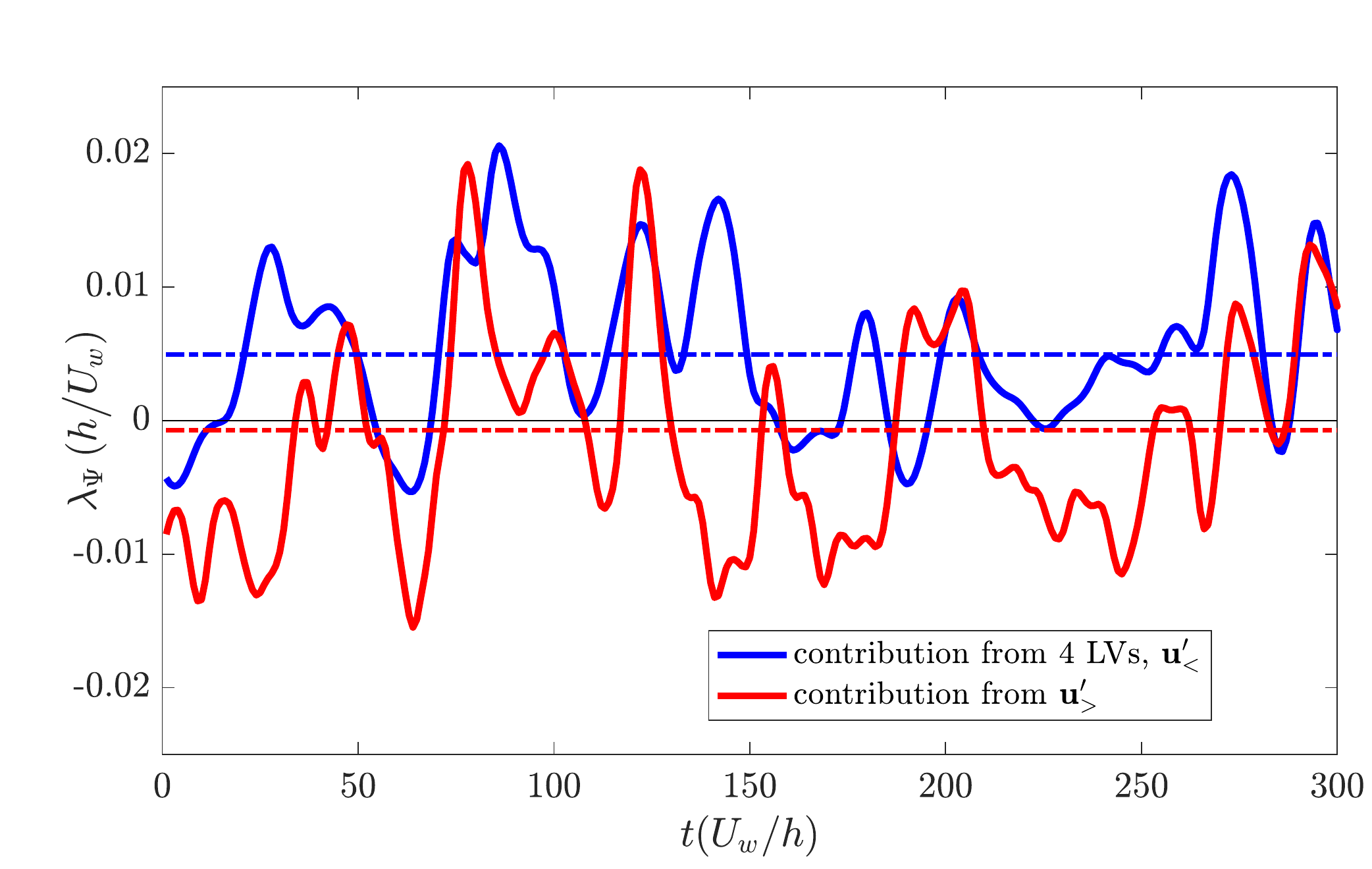}
\vspace{-2em}
\caption{Time evolution of the contribution of the torques arising from the Reynolds stresses produced by  $\u'_<$  to  maintenance of $\Psi^2$, which largely consists of the rolls (blue). The torques from $\u'_>$ (red) make no net contribution to the rolls in this 
measure. The time mean contributions are indicated with  dashed lines.
This figure identifies the perturbation subspace responsible for maintaining the roll against dissipation to be the subspace spanned by the four least stable  LVs.} 
\label{fig:Psi}
\end{figure}
 
 \section{Analysis of the contribution of the Lyapunov vectors to the self -sustaining process}

We have seen that  the perturbation structure in a DNS has significant projection
on the first LV ($11 \%$ on average) and about $20 \%$ on average on the subspace spanned by the 
four least stable LVs.  These least stable  Lyapunov vectors also 
dominate the others in the rate of energy extraction from the streamwise flow $\U(y,z,t)$.
Remarkably, they also  account fully for the forcing of the roll and 
therefore the SSP. 
In order to assess the contribution of the Lyapunov vectors to the roll forcing
consider the equation for the streamwise component  $\Omega_x = \Delta_h \Psi$ with $\Delta_h \equiv \partial_y^2+\partial_z^2$,
of the mean  vorticity equation, which is obtained by taking the streamwise component of the curl of \eqref{eq:NSm}:
\begin{equation}
\frac{D \Omega_x}{D t} =   \underbrace{- \left [ (\partial^2_{y}-\partial^2_{z} )\langle vw \rangle_x  + \partial_{yz} \left ( \langle w^2 \rangle_x - \langle v^2 \rangle_x  \right ) \right ] }_{G_{\Omega_x}} + \nu \Delta_h \Omega_x ~,
\label{eq:MPSI}
\end{equation}
where $D/Dt = \partial_t + \U \cdot \nabla$ is the substantial derivative on the streamwise mean flow. 
%\begin{align}
%\partial_t \Omega_x & = \underbrace{ -\left ( V \partial_y + W\partial_z \right ) \Omega_x}_A+ \underbrace{ \nu \Delta_h \Omega_x}_{D}\nonumber \\
%& \underbrace{- \left [ (\partial^2_{y}-\partial^2_{z} )\langle vw \rangle_x  + \partial_{yz} \left ( \langle w^2 \rangle_x - \langle v^2 \rangle_x  \right ) \right ] }_{G_{\Omega_x}} ~,
%\label{eq:MPSI}
%\end{align}
%Terms $A$ and $D$  represent advection and dissipation of $\Omega_x$ in the $(y-z)$ plane
From \eqref{eq:MPSI} we see that if  it were not for the streamwise mean torque from the perturbation Reynolds stresses,  $G_{\Omega_x}$,  the roll
would decay.
The contribution of perturbation Reynolds stresses  to the rate of change of the normalized
 streamwise square vorticity 
can be measured  by $\lambda_{\Omega_x} = {\int_V {\Omega_x} G_{\Omega_x} dV}/({2 \int_V  \Omega_x^2 dV})$,
and similarly, if more emphasis is to be given to the large scales,
we could use as a measure  the contribution of the perturbation Reynolds stresses
to the maintenance of  the square of	 the streamfunction.
This normalized measure of contribution to $\Psi^2$ is
$\lambda_{\Psi} = {\int_V {\Psi} G_{\Psi} dV}/({2 \int_V  \Psi^2 dV})$,
where $G_\Psi \equiv \Delta_h^{-1} G_{\Omega_x}$ and $\Delta_h^{-1}$ is the inverse 
 cross-stream/spanwise Laplacian.
In order to analyze the contribution of the first few Lyapunov vectors to the maintenance of the roll component of the SSP
we decompose the perturbation field $\u'$ into its 
component, $\u_{<}'$, projected on the subspace spanned by the 4 least damped energy orthonormal LVs,
denoted $\u_i'$, $i=1,2,3,4$ and the projection on the complement $\u_>'$:
\begin{equation}
\u_{<}' \equiv \sum_{i=1}^4 (\u'\cdot \u_i' ) \u_i'~,~~~\u_{>}' \equiv \u'-\u_{<}'~,
\end{equation} 
and estimate  $G_\Psi$ produced by $\u_<'$ and $\u_>'$. 
The contribution of these subspaces to  $\lambda_\Psi$ is shown  in Fig. \ref{fig:Psi}.
It can be seen that the first four least stable LVs contribute $100 \% $ on average 
to  the roll maintenance \footnote{The first LV contributes to $\lambda_\Psi$ 
on average $60\%$, while inclusion of the second LV adds another $26 \%$. 
The corresponding contribution to
$\lambda_{\Omega_x}$ by $\u_<'$ is $20 \%$ consistent with more emphasis 
being placed on small scale vorticity by the square vorticity measure.}.
 This identification of a small subset of the least stable  LVs
as the perturbation structures that support  the SSP  anticipates
laminarization of the turbulence in the DNS upon removal of this subspace (cf. \cite{Farrell-Ioannou-2018-control}).

%  vector.contained    
% span the primary subspace of the perturbation energetics and that 
% the alternative bases of the directions of 
% greatest mean and of greatest 
% instantaneous energy growth rate are inferior 
% bases for the energetics which is 
% consistent with parametric growth being 
% the mechanism supporting the perturbation variance in turbulent 
%Couette flow (????) . 
%
%At the Reynolds  number $R=600$
%used in this study there is only one such Lyapunov vector (with twofold degeneracy) so the entire support of the turbulent state is by this vector.  However, under either a stochastic parameterization for N_4 or, as in the present study, inclusion of the N_4 term in the dynamics, the perturbation energy becomes supported by a larger set of structures due to nonlinear energy scattering~\citep{Farrell-Ioannou-2017-sync}.  We are interested to determine whether the primary support of both the energy of the perturbations and the energy transfer from the mean to the perturbations continues to be by the Lyapunov vectors which would be expected if the parametric mechanism was the primary nonlinear mechanism supporting the NL turbulence as it is for the support of the RNL turbulence.

% The turbulent mean flow $\U$ is such so that
% perturbations that lead to large instantaneous growth rate have large spanwise 
% wavenumber and are located at the very high shear regions next to the
% channel boundaries. Consequently, the state has a small projection 
% on the configurations that lead to explosive growth.
    
\section{Conclusions}
\vskip0.2in

Analyses made using SSD systems closed at second order have  
 demonstrated that a realistic self-sustained turbulent state is maintained by  the parametric growth mechanism arising from 
 interaction  between the  temporally and spanwise/cross-stream spatially varying  streamwise-mean flow and 
 the associated perturbation component~\cite{Thomas-etal-2014,Thomas-etal-2015,Farrell-etal-2016-VLSM}.
  In these second order SSD systems the streamwise-mean flow 
 is necessarily adjusted   exactly to neutral stability, with the understanding that the time
 dependent streamwise mean flow is considered neutral when the first Lyapunov exponent is  zero.
 This result reinterprets the conjecture that the statistical state of inhomogeneous 
  turbulence should have mean flow  adjusted  to neutral hydrodynamic stability.

  In this work  identification of the parametric mechanism supporting the perturbation component of turbulence  
  obtained using SSD in the RNL system has been  extended to DNS.  
  While in the case of RNL support of the energy, energetics and the roll forcing is 
  solely  on the feedback neutralized first Lyapunov vector, 
  in  the case of DNS, energy, energetics and roll forcing are spread by nonlinearity over  the Lyapunov vectors. 
  Indicative that the Lyapunov vectors maintain their centrality in DNS dynamics 
  is that support 
  of the perturbation structure and energetics  is ordered in the Lyapunov vectors 
  descending in   their associated exponents.
  The neutrality of the top Lyapunov vector in both RNL and DNS, when account is taken of the transfer of energy to other 
  scales in the case of DNS,  implies that the mean state neutrality conjecture for determining the statistical state 
  is valid if neutrality of the mean state is 
  reinterpreted as neutrality of the top Lyapunov vector(s). Consistent with the parametric 
  mechanism sustaining the turbulence, 
   the perturbation structure is concentrated on
  the top Lyapunov vectors of the time varying streamwise-mean flow and ordered in their Lyapunov exponents.
  Identification  of the dynamical support of RNL and DNS turbulence to be the neutrally and stable Lyapunov 
  vectors with associated parametric growth mechanism vindicates the conjecture 
  that the mechanism that underlies turbulence in wall-bounded   shear flow  
   is parametric instability of  the time and 
   spanwise varying  streamwise mean. Although essentially unstructured 
   scattering by perturbation-perturbation nonlinearity constitutes a plausible  mechanism by which the subspace of 
   transiently growing perturbations is supported, we find    
   the perturbation-perturbation nonlinearity 
  does not configure the perturbations to extract more energy from the mean flow  
 than they would in the absence of this term implying
 that  the nonlinearity  acts as
a disruption to the parametric growth process supporting the  perturbation field
 rather than augmenting the perturbation maintenance  process.
 The perturbation-perturbation nonlinearity instead transfers 
 energy to other Lyapunov vectors maintaining them as parametric energy extracting structures 
 despite their 
 negative exponents.  These Lyapunov vectors that are being 
excited by scattering and maintained by extracting energy from the mean flow and are
 primarily responsible for the structure and maintenance of the perturbation 
 field in contradistinction to the familiar implication of a perturbation cascade in this process. 
 
  We conclude that the mean flow in the DNS has been adjusted to Lyapunov
  neutrality and that the Lyapunov vectors support the energy, energetics and role  in the SSP
  of the perturbation component of the turbulent state. These properties of the Lyapunov vectors verify  
  that parametric growth  on the fluctuating 
  streamwise mean flow and its regulation by Reynolds stress 
  feedback, which has been identified in RNL turbulence, is also the mechanism underlying 
  the support of as well as the regulation to a statistical steady state of turbulence in the DNS.

\begin{acknowledgments}
This work was funded in part by the Coturb program of the European Research Council. We thank Javier Jimenez for his useful comments and discussions.
Marios-Andreas Nikolaidis gratefully acknowledges the support of the Hellenic Foundation for Research 
and Innovation (HFRI) and the General Secretariat for Research and Technology (GSRT). 
Brian F. Farrell  was partially supported by NSF AGS-1246929.
 \end{acknowledgments}
%\bibstyle{apsrev4-1.bst}
%\bibliography{../../bibfile/basic_references_20110713}

%\bibliography{../bibfile/basic_references}
%\bibliography{../../bibfile/basic_references_20110713}


\end{document}